\input harvmac

\def\journal#1&#2(#3){\unskip, \sl #1\ \bf #2 \rm(19#3) }
\def\andjournal#1&#2(#3){\sl #1~\bf #2 \rm (19#3) }

\def\ie{{\it i.e.}}
\def\eg{{\it e.g.}}

\def\frac#1#2{{#1\over#2}}

\def\half{\frac12}

\def\inbar{\,\vrule height1.5ex width.4pt depth0pt}
\def\IC{\relax\hbox{$\inbar\kern-.3em{\rm C}$}}
\def\IR{\relax{\rm I\kern-.18em R}}
\def\IP{\relax{\rm I\kern-.18em P}}
\def\IZ{\relax{\rm I\kern-.18em Z}}

%
%

%
\catcode`\@=11
\def\slash#1{\mathord{\mathpalette\c@ncel{#1}}}
\overfullrule=0pt

\def\QQ{{\cal Q}}

\def\underrel#1\over#2{\mathrel{\mathop{\kern\z@#1}\limits_{#2}}}

\catcode`\@=12


%

\def \sinh{{\rm sinh}}


\lref\KutasovRR{
D.~Kutasov,
``Accelerating Branes and the String/Black Hole Transition,''
arXiv:hep-th/0509170.
}

\lref\GiveonRW{
  A.~Giveon, A.~Konechny, E.~Rabinovici and A.~Sever,
  ``On thermodynamical properties of some coset CFT backgrounds,''
  JHEP {\bf 0407}, 076 (2004)
  [arXiv:hep-th/0406131].
}

\lref\GiveonFU{
  A.~Giveon, M.~Porrati and E.~Rabinovici,
  ``Target space duality in string theory,''
  Phys.\ Rept.\  {\bf 244}, 77 (1994)
  [arXiv:hep-th/9401139].
}

\lref\MyersUN{
  R.~C.~Myers and M.~J.~Perry,
  ``Black Holes In Higher Dimensional Space-Times,''
  Annals Phys.\  {\bf 172}, 304 (1986).
}

\lref\GiveonMI{
  A.~Giveon, D.~Kutasov, E.~Rabinovici and A.~Sever,
  ``Phases of quantum gravity in AdS(3) and linear dilaton backgrounds,''
  Nucl.\ Phys.\ B {\bf 719}, 3 (2005)
  [arXiv:hep-th/0503121].
}

\lref\VenezianoZF{
  G.~Veneziano,
  ``A Stringy Nature Needs Just Two Constants,''
  Europhys.\ Lett.\  {\bf 2}, 199 (1986).
}

\lref\SusskindWS{
  L.~Susskind,
  ``Some speculations about black hole entropy in string theory,''
  arXiv:hep-th/9309145.
}

\lref\HorowitzNW{
  G.~T.~Horowitz and J.~Polchinski,
  ``A correspondence principle for black holes and strings,''
  Phys.\ Rev.\ D {\bf 55}, 6189 (1997)
  [arXiv:hep-th/9612146].
}

\lref\McGuiganQP{
  M.~D.~McGuigan, C.~R.~Nappi and S.~A.~Yost,
  ``Charged black holes in two-dimensional string theory,''
  Nucl.\ Phys.\ B {\bf 375}, 421 (1992)
  [arXiv:hep-th/9111038].
}

\lref\GiveonPX{
  A.~Giveon and D.~Kutasov,
  ``Little string theory in a double scaling limit,''
  JHEP {\bf 9910}, 034 (1999)
  [arXiv:hep-th/9909110].
}

\lref\GiveonYG{
  A.~Giveon, B.~Kol, A.~Ori and A.~Sever,
  ``On the resolution of the time-like singularities in Reissner-Nordstroem
  and negative-mass Schwarzschild,''
  JHEP {\bf 0408}, 014 (2004)
  [arXiv:hep-th/0401209].
}

\lref\GiveonZZ{
  A.~Giveon and A.~Sever,
  ``Strings in a 2-d extremal black hole,''
  JHEP {\bf 0502}, 065 (2005)
  [arXiv:hep-th/0412294].
}

\lref\GiveonGE{
  A.~Giveon, E.~Rabinovici and A.~Sever,
  ``Beyond the singularity of the 2-D charged black hole,''
  JHEP {\bf 0307}, 055 (2003)
  [arXiv:hep-th/0305140].
}

\lref\JohnsonJW{
  C.~V.~Johnson,
  ``Exact models of extremal dyonic 4-D black hole solutions of heterotic
  string theory,''
  Phys.\ Rev.\ D {\bf 50}, 4032 (1994)
  [arXiv:hep-th/9403192].
}

\lref\GiveonJJ{
  A.~Giveon and M.~Rocek,
  ``Generalized duality in curved string backgrounds,''
  Nucl.\ Phys.\ B {\bf 380}, 128 (1992)
  [arXiv:hep-th/9112070].
}

\lref\fzz{V. Fateev, A. B. Zamolodchikov and Al. B. Zamolodchikov,
unpublished.}

\lref\KazakovPM{
 V.~Kazakov, I.~K.~Kostov and D.~Kutasov,
 ``A matrix model for the two-dimensional black hole,''
 Nucl.\ Phys.\ B {\bf 622}, 141 (2002)
 [arXiv:hep-th/0101011].
}

\lref\PolchinskiRR{
J.~Polchinski,
``String theory. Vol. 2: Superstring theory and beyond.''
}

\lref\KutasovUA{
D.~Kutasov and N.~Seiberg,
``Noncritical Superstrings,''
Phys.\ Lett.\ B {\bf 251}, 67 (1990).
}

\lref\KarczmarekBW{
J.~L.~Karczmarek, J.~Maldacena and A.~Strominger,
``Black hole non-formation in the matrix model,''
arXiv:hep-th/0411174.
}

\lref\AtickSI{
J.~J.~Atick and E.~Witten,
``The Hagedorn Transition And The Number Of Degrees Of Freedom Of String
Theory,''
Nucl.\ Phys.\ B {\bf 310}, 291 (1988).
}

\lref\HorowitzJC{
G.~T.~Horowitz and J.~Polchinski,
 ``Self gravitating fundamental strings,''
Phys.\ Rev.\ D {\bf 57}, 2557 (1998)
[arXiv:hep-th/9707170].
}

\lref\DamourAW{
T.~Damour and G.~Veneziano,
``Self-gravitating fundamental strings and black holes,''
Nucl.\ Phys.\ B {\bf 568}, 93 (2000)
[arXiv:hep-th/9907030].
}

\Title{
\rightline{hep-th/0510211}}
{\vbox{\centerline{The Charged Black Hole/String Transition}}}
\medskip
\centerline{\it Amit Giveon${}^{1}$ and David Kutasov${}^{2}$}
\bigskip
\smallskip
\centerline{${}^{1}$Racah Institute of Physics, The Hebrew University}
\centerline{Jerusalem 91904, Israel}
\smallskip
\centerline{${}^2$EFI and Department of Physics, University of Chicago}
\centerline{5640 South Ellis Avenue, Chicago, IL 60637, USA }

\bigskip\bigskip\bigskip
\noindent

We generalize the discussion of \KutasovRR\ to charged black holes.
For the two dimensional charged black hole, which is described by
an exactly solvable worldsheet theory, a transition from the black 
hole to the string phase occurs when the Hawking temperature of the 
black hole reaches a limiting value, the temperature of free  strings 
with the same mass and charge. At this point a tachyon winding 
around Euclidean time in the Euclidean black hole geometry, which 
has a non-zero condensate, becomes massless at infinity, and the 
horizon of the black hole is infinitely smeared. For Reissner-Nordstrom 
black holes in $d\ge 4$ dimensions, the exact worldsheet CFT is not 
known, but we propose that it has similar properties. We check that 
the leading order solution is in good agreement with this proposal, 
and discuss the expected form of $\alpha'$ corrections.

\vfill
\Date{}

\newsec{Introduction}

In \KutasovRR\ a new picture of the string/black hole transition (see 
\refs{\VenezianoZF\SusskindWS\HorowitzNW-\GiveonMI} for earlier 
discussions) was proposed. According to this picture, as the Hawking 
temperature of a black hole $T_{bh}$ increases, the geometry develops 
large fluctuations in a wider and wider region around the horizon. When 
$T_{bh}$ approaches the temperature of fundamental strings with the same 
quantum numbers as the black hole, $T_f$, the size of this stretched 
horizon goes to infinity and the black hole becomes indistinguishable 
from a gas of fundamental strings. 

The divergence of the size of the stretched horizon can be seen by studying 
the Euclidean black hole geometry. It was argued in \KutasovRR\ that in this 
geometry there is a non-zero condensate of the closed string tachyon wound 
around the Euclidean time direction. This is known to be the case for two 
dimensional black holes related to $SL(2,\IR)$ CFT, and was conjectured in 
\KutasovRR\ to be the case for higher dimensional black holes as well. The 
mass of the wound tachyon at infinity is large for black holes whose Hawking 
temperature is much lower than the corresponding fundamental string temperature, 
but it goes to zero when the two temperatures coincide. In this case, the tachyon 
condensate smears the Euclidean black hole geometry all the way to infinity. It 
was argued in \KutasovRR\ that this is the Euclidean manifestation of the infinite 
size of the stretched horizon for the case $T_{bh}=T_f$. 

In \KutasovRR\ the above picture was explored and tested for the case of uncharged
black holes in two dimensional spacetime with a linear dilaton, and for Schwarzschild
black holes in $d\geq 4$ dimensions. The purpose of this note is to generalize the
discussion to charged black holes. There are a number of reasons to study this 
generalization. One is that it is known \HorowitzNW\ that the string/black hole 
transition for charged black holes can occur at temperatures that are arbitrarily low 
compared to the Hagedorn temperature. Thus, it is natural to ask whether and how 
the mass of the wound tachyon at infinity can go to zero at such temperatures. In 
particular, in the extremal limit in which the mass and charge coincide, both $T_{bh}$ 
and $T_f$ go to zero. A natural question is what happens then? Also, Wick rotating a 
charged black hole to Euclidean space leads to a background with an imaginary electric 
field, unless we  Wick rotate the charge as well. In discussing the normalizability of 
states it seems important to consider a real Euclidean background, \ie\ to Wick rotate 
the charge, but then it is not clear how to relate the results to the original Minkowski 
problem. 

The particular cases we will study are the two dimensional black hole of \McGuiganQP, 
and Reissner-Nordstrom (RN) black holes in $d\ge4$ dimensions. The former is an 
exact (in $\alpha'$) solution of string theory \refs{\GiveonJJ\JohnsonJW-\GiveonGE}. 
In section 2 we show that the Euclidean solution involves a condensate of a tachyon
which winds once around Euclidean time. This condensate becomes non-normalizable 
precisely at the point where the Hawking temperature of the black hole reaches the 
(limiting) fundamental string temperature. At the transition, the fundamental string 
and black hole entropies coincide \GiveonMI. 

In section 3, we study RN black holes in $d\ge4$ dimensions. Like in the Schwarzschild
case, the exact CFT corresponding to these geometries in string theory is not known, 
but as in \KutasovRR, we can test the picture by using the leading order solution. We
find that at the transition, when $T_{bh}=T_f$, the entropy computed from the leading 
order black hole solution differs from the fundamental string one by a factor $(d-3)/(d-2)$, 
independently of the charge. Like in \KutasovRR, we attribute this disagreement to $\alpha'$ 
corrections. In section 4 we discuss some aspects of our results.

\newsec{Charged two dimensional black hole}

\subsec{Lorentzian black hole}

The two dimensional charged black hole of \McGuiganQP\ 
is described by the line element
\eqn\rnlike{ds^2=-f(r)dt^2+{dr^2\over Q^2r^2f(r)}~,}
dilaton
\eqn\dilr{\Phi(r)=-\half\ln(r/Q)~,}
and $U(1)$ gauge field
\eqn\aatt{A_t(r)={q\over r}~.}
The function $f(r)$ in \rnlike\ is related to the mass 
$m$ and charge $q$ of the black hole as follows
\eqn\fofr{f(r)=1-{2m\over r}+{q^2\over r^2}=
\left(1-{r_-\over r}\right)\left(1-{r_+\over r}\right)~.}
The second equality in \fofr\ defines the inner $(r_-)$ and outer 
$(r_+)$ horizons of the black hole, which are given by
\eqn\rpm{r_{\pm}=m\pm\sqrt{m^2-q^2}~.}
At large $r$, $f(r)\to 1$ and the geometry approaches a linear dilaton one,
\eqn\lindil{\eqalign{
ds^2=&d\phi^2-dt^2~,\cr
\Phi=&-{Q\over2}\phi~.\cr
}}
Here and below we set $\alpha'=2$, such 
that the central charge of $\phi$ is $c_\phi=1+3Q^2$. 

The solution \rnlike\ -- \aatt\ is very reminiscent of the four
dimensional Reissner-Nordstrom black hole, and can be used as a 
toy model for studying charged black holes in higher dimensions. 
It has the advantage that string propagation in this geometry is 
described by a coset model \JohnsonJW, and is (classically) 
exactly solvable. The coset in question,
\eqn\backss{{SL(2,\IR)_k\times U(1)_L\over U(1)}~,}
can be constructed as follows (see \GiveonZZ\ for a recent discussion). 
We start with the supersymmetric $SL(2,\IR)$ WZW model at level $k$, 
whose central charge is $c=\bar c=3+{6\over k}+{3\over2}$, and add to 
it a left-moving supersymmetric $U(1)$ current with $c={3\over2}$, 
$\bar c=0$. To get the black hole \rnlike\ -- \aatt\ we gauge a $U(1)$ 
current whose right-moving component is one of the spacelike $U(1)$'s in 
$SL(2,\IR)_k$, while the left-moving component is a linear combination of a 
spacelike $U(1)$ in $SL(2,\IR)$ and the $U(1)_L$ in \backss. The free parameter 
that determines this linear combination corresponds to the charge to mass 
ratio of the black hole. When the left-moving component of the gauged 
$U(1)$ lies entirely in $SL(2,\IR)$, the coset describes the uncharged black 
hole $(q=0)$. When it is entirely in $U(1)_L$ \backss, one gets the extremal 
black hole, with $q=m$.

In the coset description, the maximal extension of the background 
\rnlike\ -- \aatt\ splits into different regions, each of which is described 
by its own set of natural coordinates. The metric, dilaton and gauge field 
in the region outside the outer horizon are given by
\eqn\metric{ds^2=d\phi^2-\left({\tanh{Q\over 2}\phi\over
1-a^2\tanh^2{Q\over 2}\phi}\right)^2d\theta^2~,}
\eqn\dilaton{\Phi(\phi)=\Phi_0-{1\over 2}\log\left(
1+(1-a^2)\sinh^2{Q\over 2}\phi\right)~,}
and
\eqn\gaugef{A_\theta(\phi)={a\tanh^2{Q\over 2}\phi\over
1-a^2\tanh^2{Q\over 2}\phi}~,}
where $a$ is a function of the charge to mass ratio of the
black hole,
\eqn\aarr{a^2={r_-\over r_+}~.}
In particular, it varies in the range $0\le |a|\le 1$. $a=0$ corresponds
to the uncharged black hole, which was studied from the current 
perspective in \KutasovRR, while $|a|=1$ corresponds to the extremal
case, $|q|=m$. The sign of $a$ can be changed by taking the gauge 
field to minus itself, \ie\ by flipping the sign of all charges. Below we will 
take $a$ and $q$ to be positive. 

The coordinates  $r$ in \rnlike\ -- \aatt\ and $\phi$ in \metric\ -- \gaugef\ 
are related by a coordinate  transformation that can be read off  \dilr, 
\dilaton, while $\theta$ and $t$ are related by
\eqn\chco{t=-{\theta\over 1-a^2}~.}
The gauge field \gaugef\ approaches a non-zero value at infinity,
\eqn\asaa{A_\theta(\phi\to\infty)\to {a\over 1-a^2}~.}
This is different from \aatt\ which goes to zero for large $r$. 
The discrepancy can be eliminated by adding the constant $-a$ to 
the right hand side of \aatt. Adding a constant to the electric potential 
does not change the classical dynamics in the black hole background. 
Nevertheless, we will see below that the presence of the non-zero 
asymptotic gauge field at infinity is important.

The coordinate $\theta$ is better suited than $t$ for studying the 
extremal limit, in which \aarr\ $a^2\to 1$ and the rescaling factor in 
\chco\ diverges. In this limit, the qualitative structure of the background 
\metric\ -- \gaugef\ changes. Instead of the asymptotically linear dilaton 
behavior \lindil, it approaches $AdS_2$ with constant dilaton,
\eqn\extback{\eqalign{ds^2=&d\phi^2-{1\over4}\sinh^2(Q\phi) d\theta^2~,\cr
\Phi=&\Phi_0~,\cr
A_\theta(\phi)=&\sinh^2{Q\over2}\phi~.\cr
}}
We will comment further on this limit in section 4. 

\subsec{Euclidean black hole and thermodynamics}

We next turn to the Euclidean continuation of the black hole solution,
which is obtained by taking $\theta\to i\theta$ in \metric. In order to 
keep the gauge field \gaugef\ real, we also need to take $a\to -ia$.
Looking back at \aarr, this can be thought of as taking $r_-\to -r_-$, 
keeping $r_+$ fixed. 

The resulting solution is the Euclidean $SL(2,\IR)\times U(1)_L\over U(1)$
coset model, recently studied in \GiveonRW. The  background fields are
\eqn\bone{ds^2=d\phi^2+R^2(\phi)dy^2~,}
\eqn\bfour{\Phi(\phi)=\Phi_0 -{1\over 2}\log\left(
1+(1+a^2)\sinh^2{Q\over 2}\phi\right)~,}
\eqn\bfive{A_y(\phi)=aR(\phi)\tanh{Q\over 2}\phi~,}
where
\eqn\btwo{R(\phi)={2\over Q}{\tanh{Q\over 2}\phi\over
1+a^2\tanh^2{Q\over 2}\phi}~.}
Regularity of the geometry at the tip, $\phi=0$, requires $y\equiv Q\theta$ 
to be periodic,
\eqn\bthree{y\sim y+2\pi~.}
Near the boundary at $\phi=\infty$, the background \bone\ -- \btwo\
approaches 
\eqn\rsu{\IR_\phi\times S^1\times U(1)_L~.}
The dilaton depends linearly on $\phi$, as in \lindil. The radius
of $S^1$ is 
\eqn\rryy{R_\infty={2\over Q(1+a^2)}~.}
The gauge field approaches 
\eqn\asgf{A_y=aR_\infty~,}
which gives rise to a non-zero Polyakov-Wilson loop
\eqn\wilinfty{
e^{i\int_0^{2\pi}dy A_y}=e^{i2\pi R_\infty a}~.}
In the coset description, $a$ is fixed by the charge to mass ratio of the 
black hole via \aarr. One can understand its value directly in gravity as 
follows. In the coordinates $(r,t)$, \rnlike, the gauge field \bfive\ is 
given by 
\eqn\gaugemod{A_t(r)={q\over r}-a~.}
The choice \aarr\ has the property that $A_t$ vanishes at the horizon,
$r=r_+$, which is a necessary condition to avoid a singularity there.

From the point of view of black hole thermodynamics, the Euclidean 
charged black hole \bone\ -- \btwo\ contributes to the canonical partition 
sum
\eqn\canpart{Z(\beta,a)={\rm Tr} e^{-\beta (H+ia\QQ)}~,}
where the inverse black hole temperature $\beta=\beta_{bh}$ and 
chemical potential $a$ are determined by the linear dilaton slope $Q$ 
and charge to mass ratio $q/m$:
\eqn\tbh{\eqalign{
(1-a^2){\beta_{bh}\over4\pi}={1\over Q}&~,\cr
a=\tan{\alpha\over 2}&~.\cr
}}
$\alpha$ parametrizes the charge to mass ratio of the black hole,
\eqn\defalpha{\sin\alpha={q\over m}~.}
Upon Wick rotation to Minkowski space, 
one finds the black hole \rnlike\ -- \aatt\
in thermal equilibrium with a heat bath with the Boltzmann factor 
\eqn\boltz{e^{-\beta (H-a\QQ)}~.}
The black hole entropy corresponding to \tbh\ can be obtained by using the 
thermodynamic relations
\eqn\tsm{\eqalign{
\beta=&\left(\partial S\over \partial m\right)_q~,\cr
-\beta a=&\left(\partial S\over \partial q\right)_m~.\cr
}}
It is given by
\eqn\entblack{S_{bh}(m,q)={2\pi\over Q}\left(m+\sqrt{m^2-q^2}\right)~.}
The partition sum associated with \boltz\ is 
\eqn\freebath{{\rm Tr} e^{-\beta (H-a\QQ)}\sim 
\int dm e^{S_{bh}(m,q)-\beta m(1-a\sin\alpha)}~.}
In principle we have to sum over all $\alpha$'s \defalpha, but we expect 
the sum to be dominated by states with a particular value of this parameter. 
Indeed, solving for $m$ in terms of the entropy and substituting into \freebath, 
we find that the integrand in \freebath\ goes like
\eqn\expfree{e^{\left(1-{Q\beta\over2\pi}f(\alpha)\right)S_{bh}}~,}
where
\eqn\ffaall{f(\alpha)={1-a\sin\alpha\over 1+\cos\alpha}~.}
For large entropy, the partition sum \freebath\ is dominated by states 
corresponding to a minimum of $f(\alpha)$. One can check that this 
minimum lies at $a=\tan{\alpha\over 2}$, which is precisely the value 
given in \tbh. This provides another way of understanding why a black 
hole with a given charge to mass ratio contributes to the partition
sum \freebath\ only for a particular value of the chemical potential. 

The value of $f(\alpha)$ \ffaall\ at the minimum is
\eqn\faamin{f(\alpha)|_{\rm min}=\half(1-a^2)~.}
Plugging \faamin\ and the black hole temperature \tbh\ into \expfree\ we 
see that the free energy vanishes, in agreement with the fact that the
density of states  with fixed $\alpha$, \entblack, exhibits Hagedorn growth.

An important property of the Euclidean black hole background is that
in addition to the fields \bone\ -- \btwo\ it has a condensate of a closed
string tachyon winding around the Euclidean time circle. For the uncharged
black hole (which corresponds to $a=0$) this is a consequence of the
generalization of the FZZ correspondence between the cigar and Sine-Liouville
theories \refs{\fzz,\KazakovPM} to the fermionic string \GiveonPX. The 
generalization to  $a\not=0$ can be obtained by performing a rotation on the
left-movers (by the angle $\alpha$ \defalpha), which mixes the $SL(2,\IR)$ and 
$U(1)_L$ factors.

At large $\phi$ the geometry \rsu\ -- \asgf\ is flat, and the vertex operator of the 
winding tachyon has the form 
\eqn\oper{T\sim e^{\tilde\beta\phi + ip_L\cdot x_L+ip_Rx_R}~.}
The left/right moving momentum vector $(p_L;p_R)$ is determined by the
radius of the $S^1$,  \rryy, and Wilson line \asgf\ (see \eg\ eq. (11.6.17) in 
\PolchinskiRR). For a state with winding one around the circle one finds
\eqn\plpr{(p_L;p_R)=\half R_\infty\left(1-a^2,2a;-(1+a^2)\right)=
{1\over Q}\left(\cos\alpha,\sin\alpha;-1\right)~.}
In the second equality we used the value of $R_\infty$ \rryy\ and the
relation between $a$ and $\alpha$  \tbh. The first component of $p_L$ 
in \plpr\ is the left-moving momentum along $S^1$, while the second 
component is in the $U(1)_L$ direction in \rsu. The last component of 
\plpr\ is the right-moving momentum on $S^1$. The length of the left 
and right moving momentum vectors is 
\eqn\pppp{p_L^2=p_R^2={1\over4}R_\infty^2(1+a^2)^2={1\over Q^2}~,}
independent of $a$. The value of $\tilde\beta$ in \oper\ is the same as in the 
uncharged case \GiveonPX, $\tilde\beta=-1/Q$. 

The fact that the tachyon \oper\ has a non-zero expectation value in the
Euclidean coset implies that the worldsheet Lagrangian contains the $N=2$ 
Liouville perturbation 
\eqn\ds{\delta S=\mu\int 
d^2z d^2\theta e^{-{1\over Q}(\phi+i(\cos\alpha,\sin\alpha)\cdot x_L-ix_R)}
+{\rm c.c.}~.}
For small $Q$, the perturbation \ds\ provides a small correction to the 
Euclidean background \bone\ -- \btwo\ since it goes rapidly to zero at large $\phi$.
As $Q$ increases, it becomes more and more important and eventually takes over. 
For $Q^2>2$, it becomes non-normalizable \KutasovUA, and the Euclidean black hole
ceases to contribute to the partition sum \canpart; see \refs{\KarczmarekBW,\GiveonMI}
for recent discussions. 

From the spacetime point of view, the asymptotic mass of the winding tachyon is a sum
of three contributions: the mass of the closed string tachyon (which is $-1$ in our units),
a contribution due to winding \pppp, and a factor $Q^2/4$ due to the linear dilaton \lindil,
\eqn\meinf{m^2_{\infty}=-1+{1\over Q^2}+{Q^2\over4}=\left({1\over Q}-{Q\over 2}\right)^2~.}
For small $Q$, the tachyon is very massive, and its wave function goes rapidly to zero \ds. 
As $Q$ increases, the tachyon becomes lighter, and modifies the geometry in a larger 
region in $\phi$. For $Q^2\to 2$, the size of this region diverges. For larger $Q$, the 
tachyon becomes massive again, but it does not have a normalizable state in the throat, 
and the background \ds\ remains non-normalizable.

\subsec{The black hole/string transition}

As is familiar from other contexts in string theory (see \eg\ \AtickSI), the presence of a 
light tachyon wound around Euclidean time for $Q^2\simeq 2$ encodes the effects of 
highly excited fundamental strings. The transition at $Q^2=2$ is from a black hole 
phase to a perturbative string one \GiveonMI. In this subsection we describe the high 
energy thermodynamics of charged perturbative strings, and the transition between
the black hole and string phases. 

The entropy of perturbative single string states with mass $m$ and charge $q$ 
under $U(1)_L$ in \rsu\ is given by
\eqn\sf{S_f(m,q)=2\pi\sqrt{1-{Q^2\over 4}}\left(m+\sqrt{m^2-q^2}\right)~.}
The corresponding  temperature is $T_f=1/\beta_f$ \tsm, with
\eqn\thag{(1-a^2){\beta_f\over4\pi}=\sqrt{1-{Q^2\over 4}}~.}
The chemical potential $a$ \tsm\ is the same as in the black hole case \tbh.
This is due to the fact that the fundamental string entropy \sf, like the black
hole entropy \entblack, is a function of the combination $m+\sqrt{m^2-q^2}$.
It is also interesting that the chemical potential $a$ only depends on $q\over m$
and not on $Q$, \ie\ it does not receive $\alpha'$ corrections. We will see that
something similar seems to happen for $d$-dimensional RN black holes. 

The temperature $T_f$ is in fact a limiting temperature for strings with
chemical potential $a$. To see that, consider the free string partition sum 
\freebath,
\eqn\freest{{\rm Tr} e^{-\beta(H-a\QQ)}\sim \int dm e^{S_f(m,q)-\beta m(1-a\sin\alpha)}~,}
where $\alpha$ is defined as in \defalpha\ again. Performing the sum \freest\ 
over states with a given $q/m$, or given $\alpha$, one finds that the integral diverges
for $\beta<\beta_c(\alpha)$, where
\eqn\betacc{\beta_c(\alpha)={2\pi\over f(\alpha)}\sqrt{1-{Q^2\over 4}}~,}
and $f(\alpha)$ is given by \ffaall. The states which lead to the smallest critical
temperature are those for which $f(\alpha)$ is smallest. The value of $\alpha$ at
the minimum was found in the previous subsection to be given by the second line
of \tbh. Plugging in
the value of $f(\alpha)$ at the minimum, \faamin, we conclude that the partition sum 
\freest\ diverges for $\beta<\beta_f$ \thag. Thus, $1/\beta_f$ is a limiting temperature 
in free string theory with the chemical potential $a$ \tbh.\foot{For the uncharged case 
$\alpha=a=0$, it reduces to the Hagedorn temperature in the linear dilaton  throat.}

Comparing \tbh\ and \thag\ we see that 
\eqn\ratiotemp{{T_{bh}\over T_f}=Q\sqrt{1-{Q^2\over 4}}~.}
For small $Q$ the black hole temperature $T_{bh}$ is much smaller than the limiting 
temperature $T_f$. As $Q$ increases, the ratio \ratiotemp\ increases until it goes to 
one at $Q^2=2$. At that point
the black hole temperature approaches the limiting one and the black hole becomes
indistinguishable from a gas of fundamental strings at the limiting temperature. Therefore,
it must be that the black hole entropy approaches the fundamental string one \KutasovRR.
This is indeed the case \GiveonMI, as can be checked by comparing \entblack\ and \sf.

To understand the transition better it is useful to compute the size of the stretched
horizon of the black hole. As in \refs{\SusskindWS,\KutasovRR}, the stretched horizon 
can be defined as the region in which the locally measured Hawking temperature 
exceeds the limiting temperature $T_f$. The local Hawking temperature is the 
temperature at infinity divided by the red-shift factor $\sqrt{g_{00}}=\sqrt{f(r)}$ 
(see \rnlike). In the coordinates \metric, it takes the form
\eqn\toft{T_{bh}(\phi)={Q\over 4\pi}{1-a^2\tanh^2{Q\over 2}\phi\over \tanh{Q\over 2}\phi}~.} 
The stretched horizon is the region in which $T_{bh}(\phi)>T_f$, or 
\eqn\comptt{{Q\over 4\pi}{1-a^2\tanh^2{Q\over 2}\phi\over \tanh{Q\over 2}\phi}>
{1\over 2\pi\sqrt{4-Q^2}}\left(1-a^2\right)~.}
For small $Q$, the size of this region is $\delta\phi\simeq 2/(1-a^2)$. As $Q$ increases,
its size grows, until it diverges for $Q=\sqrt2$. 

It is interesting to note that the mass of the wound tachyon, \meinf,
is related to the black hole and fundamental string temperatures \tbh, \thag\
by the relation
\eqn\relmbb{m_\infty^2=(1-a^2)^2\left[\left(\beta_{bh}\over4\pi\right)^2-
\left(\beta_f\over4\pi\right)^2\right]~.}
This relation is very reminiscent of the one satisfied by the thermal scalar in 
critical string theory at finite temperature  \AtickSI.  It generalizes the one that 
was obtained in \KutasovRR\ for the uncharged case $a=0$, and we propose 
the same interpretation as the one given there. In general, the thermal scalar 
encodes properties of highly excited strings, and the fact that it has a non-zero 
condensate \ds\ suggests that these strings are excited in the black hole 
background.

\newsec{Reissner-Nordstrom black holes}

The $d$-dimensional RN black hole is a solution of the equations
of motion of Einstein gravity coupled to a $U(1)$ gauge field. 
Unlike the two dimensional case, this solution is expected to 
receive $\alpha'$ corrections, which we will briefly comment on below. 

The line element is given by \MyersUN 
\eqn\drn{ds^2=-f(r)dt^2+{dr^2\over f(r)}+r^2d\Omega_{d-2}^2~,}
with
\eqn\deff{f(r)=\left(1-{r_-^{d-3}\over r^{d-3}}\right)
\left(1-{r_+^{d-3}\over r^{d-3}}\right)~.
}
There is also a gauge potential $A_t\sim q/r^{d-3}$. In comparing to string 
theory we will take  $A_\mu$ to be  a Kaluza-Klein gauge field, whose charge 
is the left-moving momentum in a compact direction.  In particular, the solutions
we will study have vanishing RR fields.

The inner and outer horizons $r_\pm$ in \deff\ are given in terms of the mass 
and charge by 
\eqn\exprpm{
r_\pm^{d-3}={8\pi G_d\over (d-2)\Omega_{d-2}}\left(m\pm\sqrt{m^2-q^2}\right)~.}
$\Omega_{d-2}$ is the area of the unit $(d-2)$-sphere, 
$\Omega_{d-2}=2\pi^{d-1\over 2}/\Gamma({d-1\over 2})$, and $G_d$ is the 
$d$-dimensional Newton constant. The charge $q$ is normalized such that
the extremal case corresponds to $q=m$. 

The Bekenstein-Hawking entropy of the black hole \drn\ is given by the
area formula,
\eqn\bhentr{S_{bh}={\Omega_{d-2}\over 4G_d}r_+^{d-2}~.}
From it we can read off the temperature and chemical potential
using \tsm. One finds
\eqn\chempot{\eqalign{
a=&\left(r_-\over r_+\right)^{d-3\over2}=\tan{\alpha\over2}~,\cr
\beta_{bh}=&{4\pi\over d-3}{r_+^{d-2}\over r_+^{d-3}-r_-^{d-3}}
={4\pi r_+\over(d-3)(1-a^2)}~,\cr}}
where $\alpha$ is defined in \defalpha. Like in the two dimensional case,
$a$ in \chempot\ can be thought of as the value of the gauge field 
at infinity. It plays a similar role in the Euclidean black hole geometry.

Note also that as a function of $q/m$ the chemical potential $a$ is the same 
in all cases (see \tbh, \chempot). In the two dimensional case the relation 
between $a$ and  $q/m$ was explained by considering the partition sum 
\freebath\ and asking which states make the biggest contribution to it. 
The same explanation can be used in the present case. Indeed, the entropy 
\bhentr\ is a function of $r_+$, which can be written in terms of $m$ and 
the parameter $\alpha$ \defalpha\ as follows:
\eqn\rplusalpha{
r_+^{d-3}={8\pi G_d\over (d-2)\Omega_{d-2}}m(1+\cos\alpha)~.}
Repeating the logic of equations \freebath\ -- \faamin\ one
finds that the largest contribution to \freebath\ for given entropy
comes from states with $\alpha$ that is related to $a$ via the  
first line of \chempot.

In the two dimensional case of section 2, we saw that the Euclidean black hole 
CFT includes a condensate of the closed string tachyon with winding one around
Euclidean time \ds. Following \KutasovRR, we postulate that a similar condensate
exists in the Euclidean Reissner-Nordstrom solution, which is obtained from the 
Lorentzian geometry \drn\ by taking $t\to it$ and $r_-^{d-3}\to -r_-^{d-3}$. 
The resulting geometry in the $(r,t)$ directions is a semi-infinite cigar 
with asymptotic radius 
\eqn\asrad{R_\infty={r_+\over d-3}{2\over 1+a^2}~.}
The tachyon wraps the Euclidean time direction, in the presence of a Wilson line
$a$ \chempot, as in section 2.

At large $r$ the vertex operator of the winding tachyon is
\eqn\windtach{T\sim {1\over r^{d-3}}e^{-k_0r + ip_L\cdot x_L+ip_Rx_R}~,}
where
\eqn\ppllpprr{(p_L;p_R)=\half R_\infty\left(1-a^2,2a;-(1+a^2)\right)=
{r_+\over d-3}\left(\cos\alpha,\sin\alpha;-1\right)~.}
$k_0$ is determined by the length of the vector \ppllpprr,
\eqn\fffg{p_L^2=p_R^2={1\over4}R_\infty^2(1+a^2)^2=\left({r_+\over d-3}\right)^2~,}
and the mass-shell condition for the tachyon,
\eqn\fffgb{k_0^2=\left(r_+\over d-3\right)^2-1=m_\infty^2~.}
When the horizon size in string units is large, the asymptotic mass of the tachyon
$m_\infty$ is large as well, and the tachyon condensate \windtach\ goes rapidly
to zero at large distances. As the horizon shrinks, the tachyon becomes lighter and
its condensate \windtach\ spreads to larger and larger $r$. At 
\eqn\rtrans{r_+=d-3~,}
the tachyon becomes massless, and the condensate ceases to be normalizable. As in the 
two dimensional case, at that point we expect the system to make a transition to a 
string phase. 

The entropy of perturbative single string states with mass $m$ and charge $q$ 
under a Kaluza-Klein gauge field $U(1)_L$  is given by the $Q\to 0$ limit of \sf,
\eqn\sfff{S_f(m,q)=2\pi\left(m+\sqrt{m^2-q^2}\right)~.}
The corresponding inverse temperature \tsm\ is 
\eqn\thaggg{\beta_f={4\pi\over 1-a^2}~.}
The chemical potential $a$ is again the same as that in \chempot. Taking the 
limit $Q\to 0$ in the discussion of subsection 2.3 one finds that $1/\beta_f$ is 
the limiting temperature for fundamental strings with chemical potential $a$. 

As before, the stretched horizon of the black hole is the region in which the local
Hawking temperature exceeds the limiting temperature $T_f$, or 
$\beta_{bh}\sqrt{f(r)}<\beta_f$:
\eqn\comparebeta{{4\pi\over 1-a^2} {r_+\over(d-3)}\sqrt{f(r)}<{4\pi\over 1-a^2}~.} 
For large black holes, $r_+\gg 1$, one finds that the stretched horizon has proper
size $\delta R\simeq 2/(1-a^2)$, which is very similar to the two dimensional result given
after eq. \comptt. As $r_+$ decreases, the size of the stretched horizon increases.
It diverges at the transition point \rtrans, where $T_{bh}$ \chempot\ reaches the
limiting temperature $T_f$. In fact, the mass of the winding tachyon $m_\infty$
\fffgb, and temperatures $T_{bh}$, $T_f$, satisfy again the relation \relmbb\ for
all dimensions $d$, masses $m$ and charges $q$. 

Following the logic of \KutasovRR\ and the discussion above, one expects that
at the transition point \rtrans\ the black hole and string entropies should agree.
Using equations  \exprpm, \bhentr\ and \sfff\ we find that in general
\eqn\ratiosss{{S_{bh}\over S_f}={r_+\over d-2}~.}
In particular, at the transition point \rtrans, we find that
\eqn\ratioss{{S_{bh}\over S_f}={d-3\over d-2}~,}
independently of the mass and charge. The black hole and string entropies are
not equal, but the black hole calculation was done using the leading order
RN solution \drn, which is expected to receive $\alpha'$ corrections. One might
hope that these corrections will shift the ratio \ratioss\ to one. 

It is interesting that the relation between the charge to mass ratio 
and the chemical potential given in the first line of \chempot\ 
is the same for all systems considered in this paper: two dimensional
and RN black holes, as well as fundamental strings in linear dilaton
and flat spacetime. If the entropy depends on the mass and charge
only via the combination  $m+\sqrt{m^2-q^2}$,
\eqn\corrss{S_{bh}(m,q)=F(m+\sqrt{m^2-q^2})~,}
the relation on the first line of \chempot\ follows for all $F$'s. Equation 
\corrss\ is certainly valid in the large mass limit $m,q\to\infty$. We expect 
it to be a property of the full, $\alpha'$ corrected, black hole solution, but 
have not proved that this is indeed the case. \foot{If \corrss\ is invalid, the 
chemical potential $a$ is not a function of the charge to mass ratio, but 
depends on both variables.}

\newsec{Discussion}

In this section we discuss two aspects of the results of \KutasovRR\ and 
this paper. One is the question of $g_s$ corrections to the properties of 
fundamental strings near the string/black hole transition point. The other
is what happens in the extremal limit $m=q$.  

One of the main results of \KutasovRR\ and this paper is that the black hole 
entropy should approach that of free fundamental strings when the Hawking 
temperature goes to that of fundamental strings with the same quantum 
numbers. One may object that the states we are matching have energies 
of order $1/g_s^2\simeq 1/G_d$, at which the fundamental strings are no longer free. 
Indeed, as discussed in \refs{\HorowitzJC,\DamourAW}, the effects of self 
gravity on individual string states are in general large near the string/black 
hole transition. 

It  is argued in \KutasovRR\ and this paper that the classical black hole 
sigma model describes an object that resembles more and more a cloud 
of free strings when the Hawking temperature goes to its limiting value, 
$T_f$. Therefore, its thermodynamics should approach that of free strings. 
The question of large $g_s$ corrections to properties of fundamental strings 
near the transition translates in the black hole sigma model to the question 
whether that model has large string loop corrections for arbitrarily small $g_s$ 
at $T_{bh}\simeq T_f$.

Near the transition, one expects this sigma model to receive large $\alpha'$ 
corrections but the $g_s$ corrections are, at least formally, arbitrarily small. 
If the string loop corrections to the classical black hole background are indeed 
small, one would conclude that the same is true for the thermal ensemble of
free strings at the temperature and chemical potential discussed above. 
Conversely, if the free string ensemble receives large corrections near the 
transition point, then the black hole sigma model must receive large quantum 
corrections there for arbitrarily small $g_s$. In any case, it seems that the 
matching between the classical black hole background and free strings must hold. 

In the discussion of charged black holes in sections 2 and 3 we focused on
the non-extremal case. It is natural to ask what happens when we take the
extremal limit $q\to m$. In this limit the black hole temperature $T_{bh}$
and the limiting fundamental string temperature $T_f$ go to zero (see 
\tbh, \thag, \chempot, \thaggg). Since the two temperatures are equal,
one might expect the black hole and string entropies to agree, and the
wound tachyon in the Euclidean geometry to be massless. 

These expectations are in apparent disagreement with the facts. In the two 
dimensional case, the mass of the tachyon at infinity is given by \meinf. It is 
independent of the charge to mass ratio of the black hole, and is large for small 
$Q$. Also, the black hole and string entropies \entblack\ and \sf\ do not agree 
for $q=m$, except at $Q^2=2$, which is the transition point for all $q$. Similarly, 
for $d$ dimensional RN black holes, the mass of the tachyon at infinity \fffgb\
is in general non-zero for $r_-=r_+$, and the black hole and string entropies
\bhentr\ and \sfff\ are in general different. 

To see how this conundrum is resolved in string theory, consider the
two dimensional case of section 2, which has the advantage of having
an exact coset description. This description gives the geometry \metric\ 
-- \gaugef. For $a\not=1$, this geometry is the same as that of \rnlike\ -- 
\aatt. For $a=1$, it changes qualitatively. In particular, the asymptotic 
behavior changes from linear dilaton to $AdS_2$ (see \extback). This 
$AdS_2$ can be thought of as the near-horizon geometry of the extremal 
black hole \rnlike. 

Thus, while for $a\not=1$, the CFT \backss\ describes the full black 
hole geometry, for $a=1$ it describes only the near-horizon region.  
This resolves the above puzzles, essentially by avoiding them. The 
linear dilaton region \lindil\  is pushed to infinity and is no longer 
part of the space \extback. Therefore, the behavior of the tachyon
condensate and the disagreement between the black hole and string 
entropies \entblack, \sf\ for generic $Q$ are unimportant since they 
refer to a region that is not part of the geometry. 

One can ask whether it is possible to compute the entropy of extremal 
black holes, which is given by \entblack\ with $q=m$, using the coset 
geometry \extback. We expect that it should be possible to do that by 
mapping string theory on the $AdS_2$ coset to a dual CFT using the 
AdS/CFT correspondence. 

In the higher dimensional RN geometries we do not have the analog of
the coset description of the exact background, but we expect the
situation to be similar. For $q<m$ we expect there to be an exact 
worldsheet CFT that describes the $\alpha'$ corrected RN geometry.
For $q=m$ such a geometry should not exist; instead, one should be
able to construct a worldsheet CFT which describes the near-horizon
geometry of the extremal black hole, $AdS_2\times S^{d-2}$, from which
one would be able to calculate the black hole entropy \bhentr.

\bigskip
\noindent{\bf Acknowledgements:}
We thank O. Lunin for discussions. This work is supported in part by the 
BSF -- American-Israel Bi-National Science Foundation. AG is also 
supported in part by the Israel Science Foundation and the EU grant 
MRTN-CT-2004-512194.  AG thanks the Theoretical Physics Division at 
CERN and the EFI at the University of Chicago for hospitality. The work 
of DK  is also supported in part by DOE grant DE-FG02-90ER40560.

\listrefs
\end

Finally, we remark that in the 2-d Schwarzchild-like case 
one finds that
\eqn\persch{\tilde m^2(\theta)=\left({\beta_{bh}(\theta)\over 4\pi}\right)^2
-\left({\beta_{f}\over 4\pi}\right)^2=
{Q^2\over 4}-1+{1\over Q^2}\tanh^2{Q\over 2}\theta~,}
for {\it any} $\theta$.
This implies, in particular, that the range in $\theta$
where $T_{bh}(\theta)$ is bigger than the Hagedorn temperature $T_f$ 
is the same range where $\tilde m^2(\theta)<0$,
as observed in \KutasovRR.
On the contrary, in the charged case this is no longer true.
The most extreme example is the extremal case ($a=1$),
for which both the black hole temperature and the temperature
of free strings with $m^2=q^2$ vanish for {\it any} $k$. 
In particular, since $T_{bh}=T_f$, we expect an infinitely 
stretched horizon in string theory for all $k$.
Intriguingly, the maximally asymmetric quotient 
${SL(2,\IR)\times U(1)_L\over U(1)}$ -- the exact CFT background
corresponding to the 2-d extremal black hole -- indeed captures 
only the near horizon $AdS_2$ limit \GiveonZZ.

Consequently, in string theory, the asymptoic value of the $(p_L;p_R)$ 
momenta in the Euclidean time and
the internal chiral fifth directions are as in eq. \plpr,
\eqn\asin{(p_L;p_R)=R_{\infty}\left((1-a^2,2a);-(1+a^2)\right)~,}
with $R_\infty$ now given by \rrn, while $a^2=r_-/r_+$ as in the 2-d case.
{}From this we find that
\eqn\pppp{p_L^2=p_R^2=r_+^2~.}
This implies that the quadratic term in the 
effective action of the winding tachyon is~\foot{We take the tachyon
to be an s-wave on the angular 2-sphere.}
\eqn\ssss{S=\int_{r_+}^\infty dr r^2
\left[f(r)(\partial_rT)^2+m^2(r)T^2\right]~,}
where (see appendix A) 
\eqn\fr{f(r)=\left(1-{r_-\over r}\right)\left(1-{r_+\over r}\right)~,}
and the  mass $m(r)$ has again two contributions, 
the usual closed string tachyon
mass and a term from winding around the circle:
\eqn\mmmm{m^2(r)=-1+R^2(r)\left(1+{A^2(r)\over f(r)}\right)^2~,}
with 
\eqn\rrrrf{R^2(r)=R^2_\infty f(r)~,}
and
\eqn\aofr{A(r)={q\over r}+a~.}
At $r\to\infty$, the mass approaches (see \asin,\pppp): 
\eqn\mto{m_\infty^2=r_+^2-1~.}
Thus, for a large black hole ($r_+\gg 1$) the tachyon mass is large 
at infinity, while the tachyon becomes massless at infinity when $r_+=1$. 

The inverse perturbative fundamental string temperature 
in the flat case is given by 
\eqn\tthag{\beta_{f}={4\pi r_+\over r_+ -r_-}~.}
This can be obtained, by using \tsm, from the entropy of free
excited fundamental strings with mass $m$ and charge $q$,
\eqn\sff{S_f=2\pi r_+~.} 
Note that \tthag\ and \sff\ are indeed equal to the flat limit 
($k\to\infty$) of \thag\ and \sf, respectively.
Comparing \tthag\ to \bbbh, we see that the temperature of large black holes 
is much smaller than the perturbative fundamental string temperature, 
but at the transition $r_+=1$
the black hole temperature is equal to the perturbative fundamental string one.
And comparing \sff\ to \ssbh, we find that at the transition the
two differ by a factor of two, as in the Schwarzschild black hole
\KutasovRR.

Remarks:

\item{1.}
For completeness, a change of coordinates,
\eqn\cofc{e^z={r-r_+\over r-r_-}~,}
gives rise to an action with a more standard kinetic term for $T$,
\eqn\sofz{S=(r_+-r_-)\int_{-\infty}^0 dz \left[
(\partial_z T)^2+\tilde m^2(z)T^2\right]~,}
and with a certain (uninspiring) effective mass $\tilde m^2(z)$. 
It obays $\tilde m^2(z)< 0$ if and only if $m^2(z)<0$.~\foot{Unlike the
2-d case ...} 

\item{2.}
The temperature measured by a stationary observer
at a fixed location $r$ is 
\eqn\tbhtbh{T_{bh}(r)={T_{bh}\over \chi(r)}~,}
where $\chi$ is the redshift factor in the black hole metric
$\chi=\sqrt{g_{00}}$. Using $g_{00}=f(r)$
and eqs. \bbbh, \tthag, one find that $T_{bh}(r)\geq T_f$
iff $f(r)\leq 1/r_+^2$. Hence, in the charged case, the range in $r$ where 
$T_{bh}(r)\geq T_f$ is different from the one where $m^2(r)<0$

\item{3.}
In the uncharged case, for which
$R_\infty=r_+=r_0$ and $A(r)=0$, we have 
\eqn\uncm{m^2(r)=-1+r_0^2f(r)} 
in \mmmm, as well as
\eqn\awas{\left({\beta_{bh}(r)\over 4\pi}\right)^2-
\left({\beta_f\over 4\pi}\right)^2=r_0^2f(r)-1~,}
and thus, in particular, $m^2(r)$ is negative precisely
when $T_{bh}(r)$ is bigger than the Hagedorn temperature \KutasovRR.
Again, in the charged case this is no longer true.

\item{4.}
Actually, we should have taken $r_-\to -r_-$ in eqs. \fr, \cofc, \sofz\
(though not in eq. \tthag) because, as discussed in section 2, 
in the continuation from Lorentzian to Euclidean time $a^2\to -a^2$
in such expressions.
Since practically it is a ``global change,'' this does not affect 
any of the results obtained above. Thus, we choose to keep things as
they are so that we can refer to equations like \fr\ both in the Lorentzian 
and the Euclidean cases, though we should recall that some continuation
of the parameters there is required when discussing the Euclidean geometry.